\documentclass[twocolumn,showpacs,superscriptaddress]{revtex4}
\usepackage{amsmath}
\usepackage{amssymb}
\usepackage{bm}
\usepackage{epsfig}
\usepackage{graphicx}
\usepackage{color}

\renewcommand{\Im}{\mathop{\rm Im}}

\begin{document}

\title{Spin waves in semiconductor microcavities}
\author{M.M. Glazov}
\affiliation{Ioffe Institute, 26 Polytechnicheskaya, St.-Petersburg 194021, Russia}
\affiliation{Spin Optics Laboratory, St. Petersburg State University, 1 Ul'anovskaya,
Peterhof, St. Petersburg 198504, Russia}
\author{A.V. Kavokin}
\affiliation{Spin Optics Laboratory, St. Petersburg State University, 1 Ul'anovskaya,
Peterhof, St. Petersburg 198504, Russia}
\affiliation{School of Physics and Astronomy, University of Southampton, SO17 1NJ
Southampton, United Kingdom}
\affiliation{Russian Quantum Center, Novaya 100, 143025 Skolkovo, Moscow Region, Russia}

\begin{abstract}
We show theoretically that a weakly interacting gas of spin-polarized
exciton-polaritons in a semiconductor microcavity supports propagation of
spin waves. The spin waves are characterised by a parabolic dispersion at
small wavevectors which is governed by the polariton-polariton interaction
constant. Due to spin-anisotropy of polariton-polariton interactions the
dispersion of spin waves depends on the orientation of the total polariton
spin. For the same reason, the frequency of homogeneous spin
precession/polariton spin resonance depends on their polarization degree.
\end{abstract}

\date{\today}

\maketitle

\emph{Introduction}. Spin waves are weakly damped harmonic oscillations of spin polarization. Predicted to appear in Fermi liquids over
50 years ago~\cite{silin:59eng} and discovered in the end of 1960s in 
metals~\cite{PhysRevLett.18.283} are among the most fascinating
manifestations of collective effects in interacting system. Later it was
understood theoretically that the spin waves can exist in a non-degenerate
electron gas~\cite{aronov:waves} as well as in atomic gases~\cite%
{bashkin81,ll1982}, and the spin waves were indeed observed in a
number of interacting gases such as Hydrogen and Helium~\cite%
{PhysRevLett.52.1508,Nacher1984,PhysRevLett.52.1810}, see Ref.~\cite%
{0038-5670-29-3-R02} for review. Spin waves were also observed in atomic
Bose gases, namely, in $^{87}$Rb vapors at temperatures of about
850 nK which exceeded the Bose-Einstein condensation temperature~\cite%
{PhysRevLett.88.070403}, see also Refs.~\cite%
{PhysRevLett.88.230403,PhysRevLett.88.230404,PhysRevLett.88.230405} where
this experiment was interpreted.

Recently, semiconductor microcavities with quantum wells sandwiched between
highly reflective mirrors have attracted a lot of interest in the solid
state and photonics communities~\cite{sanvitto_timofeev}. In these
artificial structures the strong coupling is achieved between excitons,
being material excitations in quantum wells, and photons confined between the
mirrors~\cite{microcavities}. Resulting mixed light-matter particles,
exciton-polaritons, demonstrate the Bose-Einstein statistics and may
condense at critical temperatures ranging from tens Kelvin~\cite{LeSiDang06}
till several hundreds Kelvin~\cite{Grandjean, Chen}, which exceeds by many
orders of magnitude the Bose-Einstein condensation temperature in atomic
gases. High transition temperatures and the strong coupling with
light makes semiconductor microcavities perfectly suited for benchtop studies
of collective effects of Bosons.

In typical GaAs based microcavities, exciton-polaritons may have two
spin projections onto the structure growth axis, $\pm 1$, corresponding to
right- and left- circular polarizations of photons (and spin moment of
excitons) forming polaritons. Owing to the composite nature of
exciton-polaritons, the interactions between them are strongly
spin-dependent~\cite{sanvitto_timofeev,microcavities,0268-1242-25-1-013001}.
A number of prominent spin-related phenomena both in interacting and in
noninteracting polariton systems have already been predicted and observed in
the microcavities, such as, e.g., polarization multistability~\cite%
{PhysRevLett.98.236401,Paraiso:2010fk} and optical spin Hall effect~\cite%
{kavokin05prl,Leyder:2007ve}, see Refs.~\cite%
{sanvitto_timofeev,microcavities,0268-1242-25-1-013001} for reviews.

Here we predict the existence of weakly-damped spin waves for a
non-degenerate or weakly degenerate polariton gas in a microcavity with
embedded quantum wells. We show that the system sustains the spin wave solutions, where the spin of polaritons $\bm S$ is harmonic function of the coordinate, $\bm r$, and time, $t$, $\bm S \propto \exp{(\mathrm i \bm q \bm r - \mathrm i \omega t)}$, and calculate their dispersion, $\omega \equiv \omega(\bm q)$. The stability of spin waves is analyzed. The experimental manifestations
of spin waves in the photoluminescence spectroscopy and spin noise studies
are discussed. {Due to the strong light-matter interaction in microcavities, which allows one to observe directly the spin states of quasi-particles, microcavities may become one of the most suitable systems for spin waves experimental studies.}

\begin{figure}[t]
\includegraphics[width=0.9\linewidth]{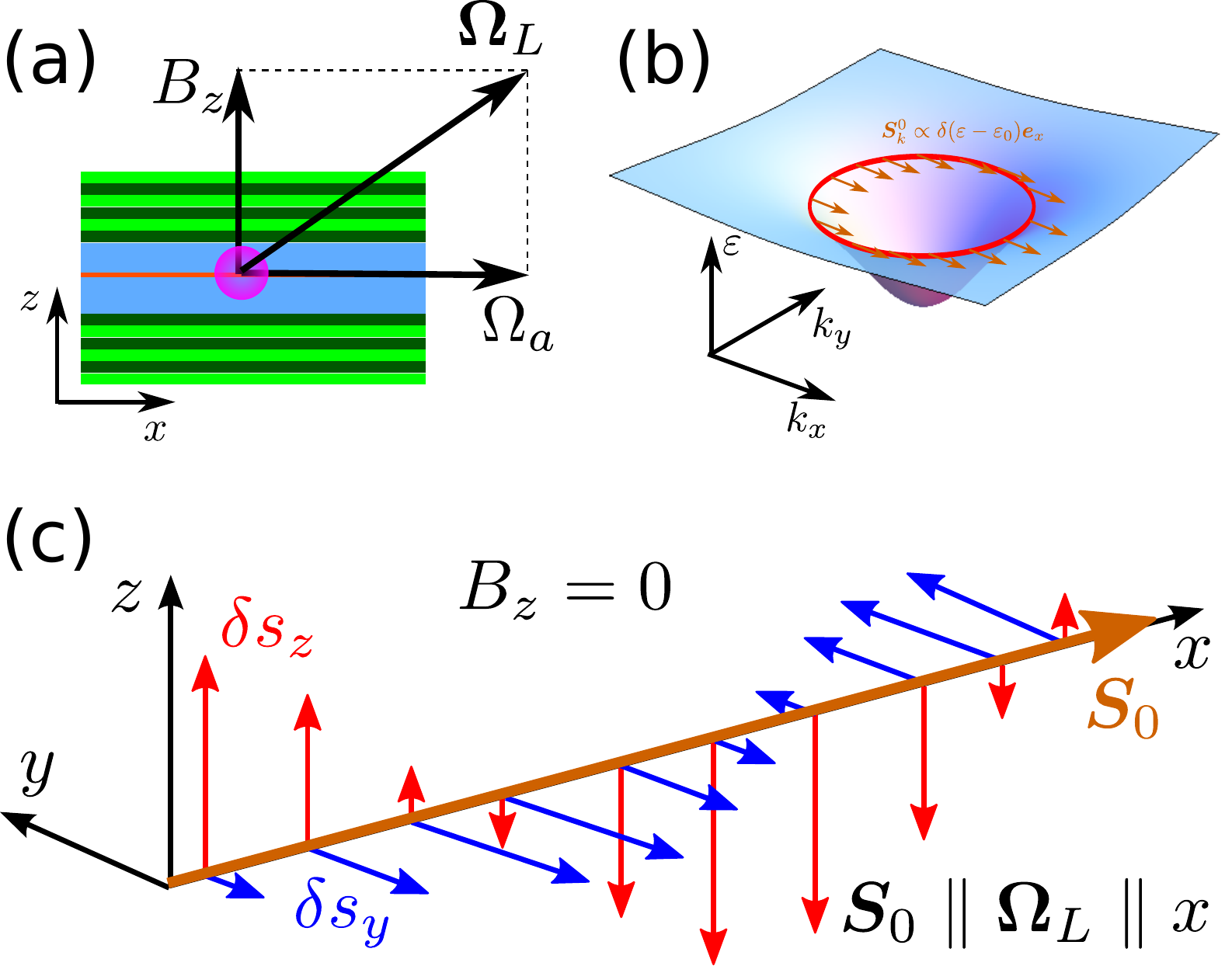}
\caption{{(a) Schematics of the microcavity structure and external fields acting on the polariton pseudospin.} (b) Sketch of polariton dispersion (surface)
and elastic circle (red
circle). The distribution function of polariton
spin in the case of
linearly polarized excitation of the states on elastic
circle is
illustrated by arrows. (c) Schematic illustration of the spin
wave propagating along $x$-axis {at $B_z=0$}; red and blue arrows demonstrate $\delta
s_z$ and $\delta s_y$ components.}
\label{fig:scheme}
\end{figure}

\emph{Model}. We consider non-degenerate or weakly degenerate polariton gas
at a temperature higher than the Berezinskii-Kosterlitz-Thouless transition
temperature {with weak interactions, in this case the} single-particle spin density matrix is parametrized as $\hat{\rho}_{\bm k}=(N_{\bm k}/2)\hat{I}+\bm S_{\bm k}\cdot 
\hat{\bm\sigma }$, where $N_{\bm k}$ is the occupancy of the orbital state
with the wavevector $\bm k$, $\bm S_{\bm k}\equiv \bm S_{\bm k}(\bm r)$ is
the coordinate $\bm r$-dependent spin distribution function, $\hat{I}$ and $%
\hat{\bm\sigma }$ are $2\times 2$ unit and Pauli matrices, respectively. The spin distribution function satisfies the kinetic
equation, which describe the rate of $\bm S_{\bm k}$ change in time as a result of the particle propagation with the group velocity $\bm v_{\bm k}$, spin precession in the effective field $\bm \Omega_{\bm k}^{(\mathrm{eff})}$, as well as generation and scattering processes described by the collision integral $\bm Q\{\bm S_{\bm k} \}$~\cite{glazov_sns_pol}:
\begin{equation}
\frac{\partial \bm S_{\bm k}}{\partial t}+\bm v_{\bm k}\cdot \frac{\partial %
\bm S_{\bm k}}{\partial \bm r}+\bm S_{\bm k}\times \bm\Omega _{\bm k}^{(%
\mathrm{eff)}}=\bm Q\{\bm S_{\bm k}\},  \label{kinetic:full}
\end{equation}%
Here
\begin{equation}
\bm\Omega _{\bm k}^{(\mathrm{eff)}}=\alpha _{1}\sum_{\bm %
k^{\prime }}S_{\bm k^{\prime },z}\bm e_{z} + \bm\Omega _{L},  \label{field}
\end{equation}%
the constant $\alpha _{1}$ describes interaction of
polaritons with parallel spins. We recall that the polariton-polariton
interactions are strongly spin-anisotropic and neglect weak interaction of particles with opposite signs of circular polarization~\cite%
{kavokin_inv,PhysRevB.82.075301,PhysRevB.77.075320,gen}. 
The effective field $\bm\Omega_{L}=\hbar^{-1}g\mu _{B}B_{z}\bm e_{z}+\Omega _{a}\bm e_{x}+\bm\Omega (\bm k)$, where $\bm e_{i}$ are unit vectors of Cartesian axes, $%
i=x,y,z$;  is interaction-independent, generally, it is contributed by
an external magnetic field $\bm B$, $g$ is the  exciton-polariton $g$-factor~\cite{Zeeman}, the splitting of linearly polarized
polariton states due to the structure anisotropy, $\Omega_a$ (the anisotropy field is parallel to $x$-axis) and TE-TM splitting of the
cavity modes, $\bm\Omega (\bm k)$. We assume that the structure anisotropy is strong enough, $\Omega_a \gg \Omega(\bm k)$ for the relevant wavevectors range and neglect TE-TM splitting~\cite{TETM}.  Under our assumptions the effective field is independent on $\bm k$ and acts similarly to the real magnetic field. Depending on the ratio of $g\mu_B B_z/\hbar$ and $\Omega_a$ this field can be arbitrarily oriented, see Fig.~\ref{fig:scheme}(a). The collision integral in the
right hand side of Eq.~\eqref{kinetic:full}
accounts for the polariton generation, scattering and decay processes, 
\begin{equation}
\bm Q\{\bm S_{\bm k}\}=-\frac{\bm S_{\bm k}}{\tau _{0}}+\bm g_{\bm k}+\sum_{%
\bm k^{\prime }}[W_{\bm k\bm k^{\prime }}\bm S_{\bm k^{\prime }}-W_{\bm %
k^{\prime }\bm k}\bm S_{\bm k}].  \label{QSk}
\end{equation}%
Here $\tau _{0}$ is the lifetime of polaritons, $\bm g_{\bm k}$ is the
polariton generation rate accounting for the in-coming flow of
quasiparticles from the reservoir, and $W_{\bm k\bm k^{\prime }}$ is the
scattering rate from the state $\bm k$ to the state $\bm k^{\prime }$ which
accounts for both elastic and inelastic scattering processes. Due to the
bosonic nature of exciton-polaritons and polariton-polariton interactions $%
\bm g_{\bm k}$ and $W_{\bm k\bm k^{\prime }}$ depend, generally, on the
occupancies and spin polarizations in the states $\bm k$, $\bm k^{\prime }$~%
\cite{glazov05a,PhysRevB.77.075320,glazov_sns_pol}. The dynamics of
polaritons can be described by Eq.~\eqref{kinetic:full} which is valid
provided that the renormalization of spectrum due to polariton-polariton
interactions is negligible, otherwise the excitations spectrum should be
found from the spin-dependent Gross-Pitaevskii equation~\cite{shelykh:066402,Hugo,Kamch}.

Under the steady-state excitation, the quasi-equilibrium distribution $\bm %
S_{k}^{(0)}$ of exciton-polaritons is formed {whose shape is determined by the generation, thermalization and interactions}. This function satisfies Eq.~%
\eqref{kinetic:full} with derivatives $\partial /\partial t$, $\partial
/\partial \bm r$ being equal to zero. For simplicity we assume that the
pumping is isotropic, hence $\bm S_{k}^{(0)}$ depends only on the absolute
value of the polariton wavevector $k=|\bm k|$. 
{Its specific form is determined by the pumping conditions. In what follows two important limiting cases are addressed: (i) quasi-resonant pumping which creates monoenergetic polaritons (so-called excitation of elastic circle), such a situation can be experimentally realized if the pump energy is slightly above the inflection point on the dispersion Fig.~\ref{fig:scheme}(b), so that the phonon-assisted relaxation towards the ground state is suppressed due to polaritons strong energy dispersion (bottleneck effect), and (ii) nonresonant pumping which creates thermalized distribution of particles.} 
{In the case of quasi-resonant excitation of
polaritons with the same energy $\varepsilon
_{0}$ by a polarized light, the spin distribution function $\bm S_{k}^{(0)}\propto \delta (\varepsilon -\varepsilon _{0})\bm e_{i}$, where $i=x$ or $y$ for the linearly polarized excitation and $i=z$ for the
circularly polarized one, $\varepsilon \equiv \varepsilon_{k}$ is the
polariton dispersion, see scheme in Fig.~\ref{fig:scheme}(b), while for thermalized polaritons (at the temperature $T$ higher than the degeneracy temperature), $\bm S_k^{(0)} \propto \bm S_0 \exp{(-\varepsilon/T)}$ with the prefactor $\bm S_0$ dependent on the effective field $\bm \Omega_L$.} In order to
analyze the spin excitations, the total spin distribution function is
presented as a sum of its quasi-equilibrium part $\bm S_{k}^{(0)}$ and the
fluctuating correction $\delta \bm s_{\bm k}\ll S_{\bm k}^{(0)}$. A standard
linearization of Eq.~\eqref{kinetic:full} and substitution of $\delta \bm s_{%
\bm k}=\exp {(\mathrm{i}\bm q\bm r-\mathrm{i}\omega t)}\bm s_{\bm k}$ with $%
\bm q$ being the wavevector and $\omega $ being the frequency of the
fluctuation yields, cf.~\cite{silin:59eng,0038-5670-29-3-R02}: 
\begin{multline}
\left[ \tau _{c}^{-1}-\mathrm{i}\omega +\mathrm{i}(\bm q\cdot \bm v_{\bm k})%
\right] \bm s_{\bm k}+\alpha _{1}\bm S_{k}^{(0)}\times \bm e_{z}\sum_{\bm %
k^{\prime }}s_{\bm k^{\prime },z}  \label{kinetic:fluct} \\
+\bm s_{\bm k}\times \bm\left( \bm\Omega _{L}+\alpha _{1}\bm e_{z}\sum_{\bm %
k^{\prime }}S_{k^{\prime },z}^{(0)}\right) =-\frac{\bm s_{\bm k}-\bar{\bm s}%
_{\bm k}}{\tau },
\end{multline}%
where we introduced the lifetime of the fluctuation $\tau _{c}$ and the
isotropization time $\tau $. Bar simbolizes averaging over possible
orientations of $\bm k$. In the simplest approximation, polaritons are
assumed to be supplied directly by the polarized pump or from the incoherent
but spin-polarized reservoir, the polariton-polariton scattering is
neglected as well as inelastic processes, and the elastic scattering is
assumed to be isotropic, in which case $W_{\bm k,\bm k^{\prime
}}=W(\varepsilon _{k})\delta (\varepsilon _{k}-\varepsilon _{k^{\prime }})$, $\tau^{-1}=\sum_{\bm k^{\prime }}W(\varepsilon _{k})\delta (\varepsilon
_{k}-\varepsilon _{k^{\prime }})$.
The lifetime of a fluctuation is governed by an interplay of polariton decay
processes accounted for by the lifetime $\tau _{0}$ in our formalism, and by
the bosonic stimulation effect, which increases the lifetime of the
fluctuations, $\tau _{c}=\tau _{0}(1+N_{k})$~\cite{glazov_sns_pol}. Equation~%
\eqref{kinetic:fluct} determines the dynamics of spin fluctuations in the
system. Its eigenmodes represent the spin waves in the interacting polariton
ensemble.

\emph{Results}. The solution of Eq.~\eqref{kinetic:fluct} can be expressed
by decomposing the function $\bm s_{\bm k}$ in the angular harmonics of the
polariton wavevector $\bm k$ as $\bm s_{\bm k}=\sum_{m}\exp {(\mathrm{i}%
m\varphi )}\bm s_{m}(\varepsilon )$, with $\varphi $ being the azimuthal
angle of $\bm k$, and reducing Eq.~\eqref{kinetic:fluct} to a system of
equations for the energy dependent functions $\bm s_{m}(\varepsilon )$. The
condition of compatibility for this system of equations yields dispersions
of the waves. Below we analyze the spectrum of excitations and eigenmodes
for different particular cases.

\emph{Homogeneous excitations}. We start the analysis from the homogeneous
case, $\bm q=0$. In this case the angular harmonics $\exp {(\mathrm{i}%
m\varphi )}\bm s_{m}(\varepsilon )$ are the eigensolutions of Eq.~%
\eqref{kinetic:fluct}. For all $m\neq 0$ one eigenmode corresponds to the
damped solution with $\bm s_{m}$ parallel to the total field $\bm\Omega ^{(%
\mathrm{tot)}}=\bm\Omega _{L}+\alpha _{1}S_{0,z}\bm e_{z}$, $\bm S_{0}=\sum_{%
\bm k^{\prime }}\bm S_{k^{\prime }}^{0}$, whose damping rate is $\nu =\tau
_{c}^{-1}+\tau ^{-1}$, and two other eigenmodes precessing in the plane
perpendicular to $\bm\Omega ^{(\mathrm{tot)}}$ with frequencies $\Omega ^{(%
\mathrm{tot)}}$ and the damping rate $\nu $.

The harmonic with $m=0$ is isotropic in $\bm k$-space, its eigenfrequency
corresponds to the spin resonance frequency. We introduce $\tilde{\bm s}%
_{0}=\sum_{\bm k}\bm s_{0}(\varepsilon )$ and perform the summation of Eq.~%
\eqref{kinetic:fluct} over $\bm k$ which yields 
\begin{equation}
(\tau _{c}^{-1}-\mathrm{i}\omega )\tilde{\bm s}_{0}-\alpha _{1}\tilde{s}%
_{0,z}\bm e_{z}\times \bm S_{0}+\alpha _{1}\tilde{\bm s}_{0}\times \bm %
e_{z}S_{0,z}+\tilde{\bm s}_{0}\times \bm\Omega _{L}=0.  \label{hom:0}
\end{equation}%
We recall that in the case of spin-anisotropic interactions the Larmor
theorem~\cite{PhysRev.67.260} in not applicable, and the homogeneous spin
excitation frequency can be renormalized by the interactions. {To illustrate it we consider the dependence of the spin resonance frequency on the orientation of effective Larmor field $\bm \Omega_L$ and spin polarization $\bm S(0)$. The orientation of $\bm \Omega_L$ in $(xz)$-plane can be varied by changing the external magnetic field contributing to $\Omega_{L,z}$ or mechanical strain contributing to $\Omega_{L,x}=\Omega_a$. We assume efficient thermalization in the spin space, $\bm S_{0}\parallel \bm\Omega _{L}$, introduce the angle $\theta$ between $\bm S_0$ and $z$-axis and present the complex
eigenfrequencies of Eq.~\eqref{hom:0} in a form}~\cite%
{gen} 
\begin{align}
& \omega _{0}=-\frac{\mathrm{i}}{\tau _{c}},  \label{hom:0:omega} \\
& \omega _{\pm }=-\frac{\mathrm{i}}{\tau _{c}}\pm \sqrt{\Omega
_{L}^{2}+\alpha _{1}^{2}S_{0}^{2}\cos ^{2}{\theta }+\alpha _{1}\Omega
_{L}S_{0}(3\cos ^{2}{\theta }-1)}.  \notag
\end{align}%
For instance, if $\bm\Omega _{L}$ and $\bm S_{0}$ are parallel to $%
z$-axis, $\theta =0$, the frequencies of the precessing modes are $\pm
|\Omega _{L,z}+\alpha _{1}S_{0,z}|$ and the damping rate is $1/\tau _{c}$.

\begin{figure}[t]
\includegraphics[width=0.85\linewidth]{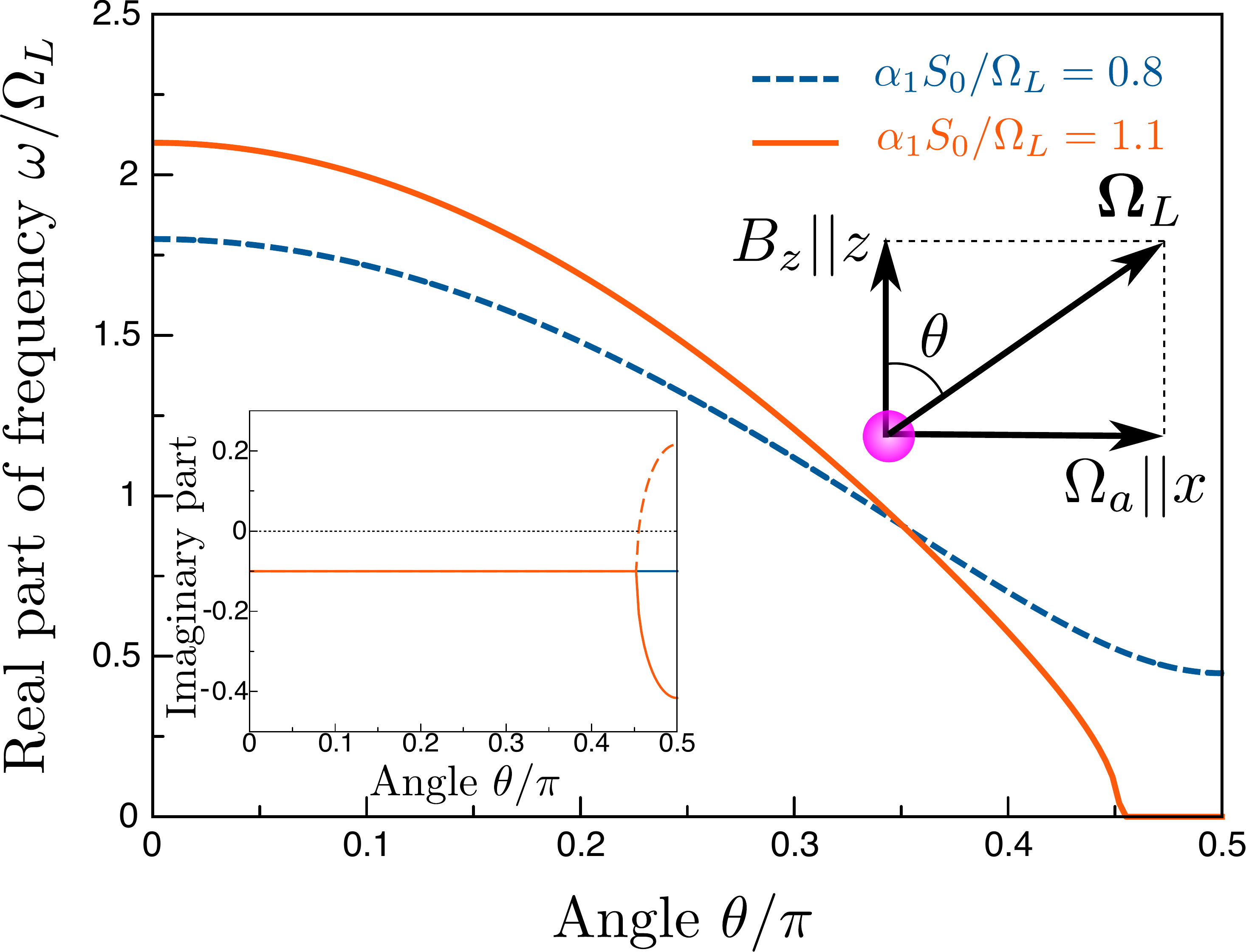}
\caption{Eigenfrequencies of homogeneous precessing modes with $m=0$ as a
function of angle $\protect\theta$ calculated after Eq.~\eqref{hom:0:omega}.
Main panel shows real part of the frequencies and inset shows imaginary
parts. The parameters of calculation are: $\protect\alpha_1 S_0\Omega_L=0.8$
(blue/dashed) and $1.1$ (red/solid), $\Omega_L\protect\tau_c=10$. }
\label{fig:m0}
\end{figure}

Real part of $\omega _{+}$ and imaginary parts of $\omega _{\pm }$ are shown
in Fig.~\ref{fig:m0} as a function of angle between the field and the $z$%
-axis $\theta $. {Depending on the sign $\alpha_1 S_0 \Omega_L$ the frequency can increase or decrease with an increase of $\theta$. Note, that for large enough $\alpha _{1}S_{0}$, hence $\omega _{\pm }$ become imaginary as shown by red/solid
curve in Fig.~\ref{fig:m0}.} Moreover, the imaginary part of one of the
frequencies can be positive which manifests the instability of the system,
see inset in Fig.~\ref{fig:m0}. To analyze it in more detail we put $%
\theta =\pi /2$ ({$B_z=0$}, $\bm S_{0}$ and $\bm\Omega _{L}$ are in the structure
plane). In this case $\omega _{\pm }=-\mathrm{i}/\tau _{c}\pm \sqrt{\Omega
_{L}^{2}-\alpha _{1}S_{0}\Omega _{L}}$. The system becomes unstable for 
\begin{equation}
\alpha _{1}S_{0}\Omega _{L}>\Omega _{L}^{2}+\frac{1}{\tau _{c}^{2}},
\label{instability}
\end{equation}%
and small fluctuations of $s_{y}$ and $s_{z}$ grow exponentially. This is
because the anisotropic interactions between polaritons favor in-plane
orientation of the pseudospin~\cite%
{0268-1242-25-1-013001,PhysRevLett.98.236401}.   
The instability of small spin fluctuations can result in the nonlinear
oscillations of spin polarization{~\cite{suppl}} similar to those discussed in Refs.~\cite%
{PhysRevB.80.195309,Kammann:2012uq} or in changes in the polarisation of the
ground state accompanied by change of orientation of $\bm S_0$ (cf.~\cite{PhysRevB.86.035310}) where the condition~\eqref{instability} no longer holds. 

\emph{Spin waves}. Let us consider the spatially inhomogeneous solutions of
Eq.~\eqref{kinetic:fluct} which describe the propagation of spin
fluctuations and spin waves. To be specific, we consider the case where $\bm %
S_{k}^{0}$ and $\bm\Omega _{L}$ are parallel to $x$-axis, and to simplify
the treatment we assume $\tau \gg \tau _{c}$~\cite{gen}. Moreover, we assume
that the system is stable at $\bm q=0$, i.e. the condition~%
\eqref{instability} is not fulfilled. We seek the solution, which
corresponds to the precessing mode at $\bm q=0$, where $s_{\bm k,x}=0$. From
Eq.~\eqref{kinetic:fluct} we arrive to the set of linear homogeneous
integral equations for $s_{\bm k,y}$ and $s_{\bm k,z}$, whose
self-consistency requirement yields 
\begin{equation}
\sum_{\bm k}\frac{\alpha _{1}\Omega _{L}\tau _{c}^{2}S_{k}^{(0)}}{[1-\mathrm{%
i}\omega \tau _{c}+\mathrm{i}(\bm q\bm v_{\bm k})\tau _{c}]^{2}+(\Omega
_{L}\tau _{c})^{2}}=1.  \label{disper:gen:x}
\end{equation}%
This equation describes the dispersion of spin waves. It has a more complex
form compared with the dispersion equation for the spin waves in the systems
with spin-isotropic interactions~\cite%
{silin:59eng,aronov:waves,0038-5670-29-3-R02}.

\begin{figure}[t]
\includegraphics[width=0.85\linewidth]{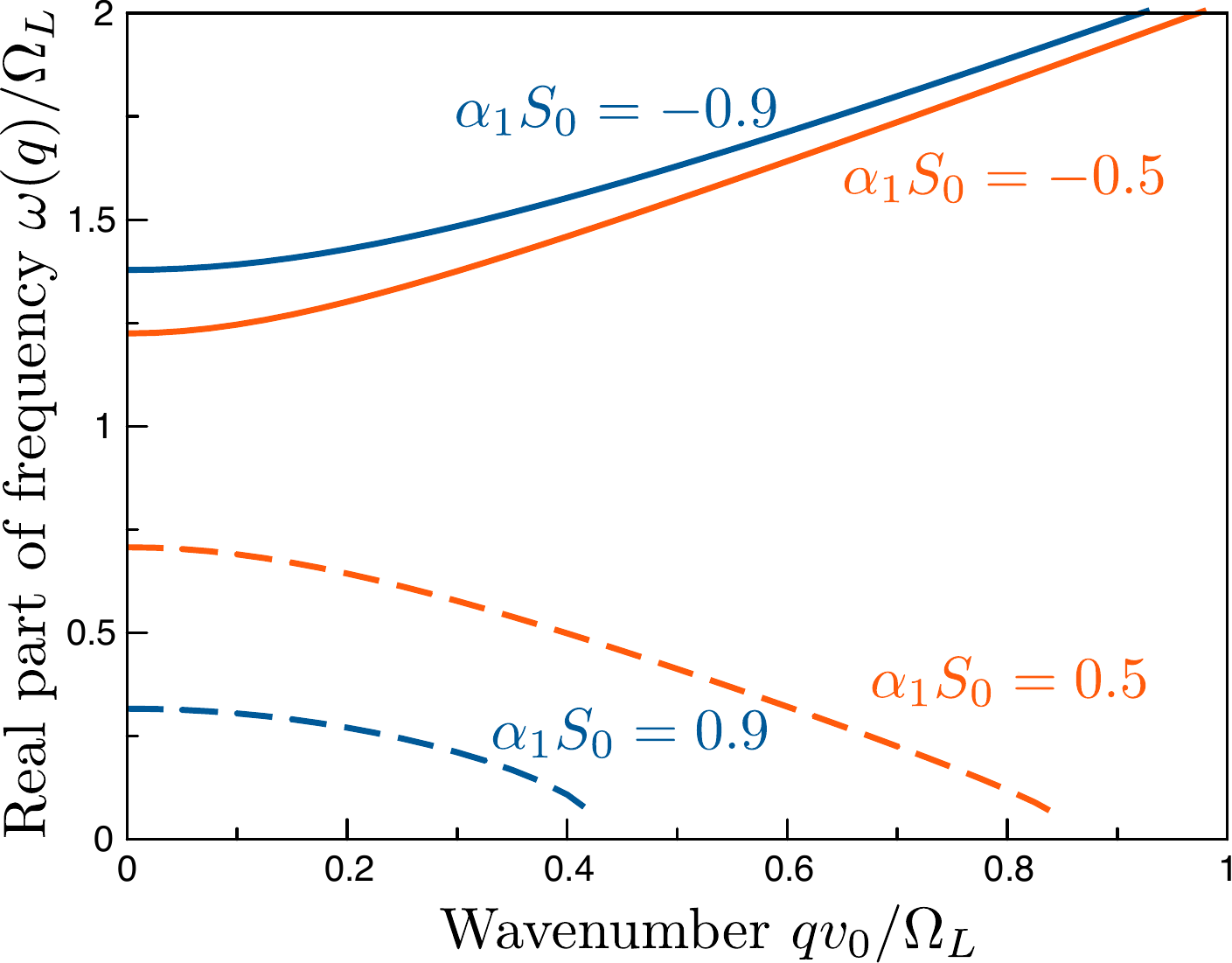}
\caption{Dispersion of spin waves in the case of resonant excitation
calculated after Eq.~\eqref{disp:el:x}. The parameters of calculation are
indicated at each curve.}
\label{fig:disper}
\end{figure}

To solve Eq.~\eqref{disper:gen:x} one has to specify the function $%
S_{k}^{(0)}$ whose form is determined by the excitation conditions. It is
instructive to consider the case of resonant excitation of polaritons at the
elastic circle where a monoenergetic distribution of particles is generated.
For the isotropic dispersion the product $\bm q\bm v_{\bm k}=qv_{0}\cos {%
\varphi }$, $v_{0}=\hbar ^{-1}d\varepsilon _{k}/dk$ is the polariton
velocity on the elastic circle. Here the summation over $\bm k$ reduces to
the averaging over the azimuthal angle $\varphi $ and Eq.~%
\eqref{disper:gen:x} takes the form
\begin{equation}
{\frac{1}{\sqrt{\tilde{\omega}_{+}^{2}-(qv_{0})^{2}}}-%
\frac{1}{\sqrt{\tilde{\omega}_{-}^{2}-(qv_{0})^{2}}}=\frac{2}{\alpha
_{1}S_{0}}},  \label{disp:el:x}
\end{equation}%
where $\tilde{\omega}_{\pm }=\omega \pm \Omega _{L}-\mathrm{i}/\tau _{c}$.
Equation~\eqref{disp:el:x} determines the dispersion of the spin waves. For $%
q=0$ it passes to $\omega _{\pm }$ in Eq.~\eqref{hom:0:omega}. For small $%
qv_{0}\ll |\alpha _{1}S_{0}|$ and $|\alpha _{1}S_{0}|\ll |\Omega _{L}|$ the
dispersion of the spin waves reads 
\begin{equation}
\omega (q)=\Omega _{L}-\alpha _{1}S_{0}/2-(qv_{0})^{2}/(\alpha _{1}S_{0})-%
\mathrm{i}/\tau _{c},  \label{disp:x:small}
\end{equation}%
where we took the solution which passes to $\omega _{+}$ in Eq.~%
\eqref{hom:0:omega}. As follows from Eq.~\eqref{disp:x:small} the dispersion
is parabolic for small $qv_{0}$ and the ``effective
mass'' is proportional to $\alpha _{1}$. Similarly to the
previously studied electronic and atomic systems \cite%
{silin:59eng,aronov:waves,0038-5670-29-3-R02,lswaves:imp} the dispersion of
spin wave results from an interplay of the gradient, $\propto \bm q\bm v_{\bm
k}$, and interaction, $\propto \alpha _{1}$, terms in the kinetic equation~%
\eqref{kinetic:fluct}. Indeed, owing to the gradient contribution, the spin
density in $\bm k$-space acquires $\propto (qv_{0})^{2}$ correction,
yielding gain or loss of energy depending on the sign of $\alpha _{1}S_{0}$.
The spin wave frequency increases with the increase of the wavevector for $%
\alpha _{1}S_{0}<0$ and decrease for $\alpha _{1}S_{0}>0$. The behavior of $%
\omega (q)$ is illustrated in Fig.~\ref{fig:disper}. Noteworthy, for the
solutions with $\alpha _{1}S_{0}>0,$ the real part of the frequency vanishes
at some $q$, in which case the solutions of Eq.~\eqref{disp:x:small} may
become unstable. For arbitrary direction of $\bm \Omega_L$ and $\bm S_0$ the dispersions of waves has a similar form, but its parameters depend on the orientation of the magnetic field and the total spin due to anisotropy of polariton-polariton interactions

It is instructive to compare the parabolic dispersion of spin waves in a
weakly interacting polariton gas with the dispersion of excitations of an
interacting polariton condensate~\cite{shelykh:066402,Hugo,Kamch}. 
In the case of a condensate of polaritons
(in the absence of TE-TM splitting) the dispersion of excitations is linear. By contrast,
the dispersion of spin waves for non-condensed polaritons is parabolic at small wavevectors.

In the case of a non-resonant excitation where a continuous distribution $%
S_{k}^{(0)}$ is formed, the analysis of the dispersion of spin waves is more
complex, particularly because the additional channel of damping caused by
the spatial dispersion appears~\cite{aronov:waves,0038-5670-29-3-R02}, but the basic physics remains the same. To illustrate this we consider the
case of a thermalized non-degenerate gas where $S_{k}^{(0)}$ is described by
the Boltzmann function characterised by an effective temperature $T$. The
evaluation of the sum in Eq.~\eqref{disper:gen:x} under the assumptions $|
\alpha_{1}S_{0}|\ll |\Omega _{L}|$, $|\alpha _{1}S_{0}|\tau _{c}\gg 1$ and $%
qv_{T}\ll |\alpha _{2}S_{0}|$, where $v_{T}=\sqrt{k_{B}T/m}$ is the thermal
velocity with $m$ being the polariton effective mass and $k_{B}$ being the
Boltzmann constant, and the solution of the resulting equation yields the
dispersion in the form 
\begin{equation}
\omega (q)=\Omega _{L}-\alpha _{1}S_{0}/2-\frac{2(qv_{T})^{2}}{\alpha
_{1}S_{0}}-\mathrm{i}\gamma _{L},  \label{disp:x:0}
\end{equation}%
where the Landau damping can be estimated as 
\begin{equation}
\gamma _{L}=\sqrt{\frac{\pi }{2}}\frac{(\alpha _{1}S_{0})^{2}}{4qv_{T}}\exp {%
\left( -\frac{(\alpha _{1}S_{0})^{2}}{8(qv_{T})^{2}}-1\right) },
\label{Landau}
\end{equation}%
and it is exponentially small for small wavevectors, in agreement with Refs.~%
\cite{aronov:waves,0038-5670-29-3-R02}. The allowance
for Landau damping in Eq.~\eqref{disp:x:0} is correct only if the damping is
large enough compared with $1/\tau _{c}$ but small compared to $|\alpha
_{1}S_{0}|$.

\emph{Conclusions}. To conclude, we predicted the existence of
exciton-polariton spin waves in semiconductor microcavities with embedded
quantum wells. The dispersion and damping of spin waves were calculated in
two important particular cases: (i) resonant excitation of a
quasi-monoenergetic distribution of polaritons at an elastic circle and (ii)
nonresonant excitation, where the Boltzmann distribution of
quasi-particles is formed. In the state-of-the-art microcavities the
polariton polarisation splittings induced by the cavity anisotropy, $\hbar
\Omega _{L}$, and interaction-induced effective field $\hbar \alpha _{1}S_{0}
$ are of the order of $100$~$\mu $eV, usually ~\cite{PhysRevLett.105.256401,PhysRevLett.106.257401,PhysRevB.79.115325}.
For the polariton lifetime $\gtrsim 10$~ps the spin waves can be readily
detectable even at the relatively weak pump. The spin waves can be excited,
e.g., in 
two-beam photoluminescence experiments where the \emph{cw} beam creates the desired
steady distribution of polaritons with a given spin polarization $\bm S_{k}^{(0)}$ and the probe beam injects a small non-equilibrium portion of
polaritons with the spin polarization different from $\bm S_{k}^{(0)}$. The
time-resolved micro-photoluminescence spectroscopy as used e.g. in Refs.~\cite%
{Lagoudakis:2008bh,Amo2009} would be a suitable tool for detection of the
spin waves. Another possibility to observe the spin waves is to use the spin
noise spectroscopy~\cite{Zapasskii:13,Oestreich:rev} and measure temporal
and spatial correlations of spin fluctuations in the presence of the pump
only~\cite{suppl}. 

\emph{Acknowledgements}. We are grateful to V.A. Zyuzin for discussions. This work was  supported by RFBR, RF President grant MD-5726.2015.2, Dynasty Foundation, Russian Ministry
of Education and Science (Contract No. 11.G34.31.0067 with SPbSU and leading
scientist A. V. Kavokin), and SPbSU grant 11.38.277.2014. 


\section*{Supplemental materials}

\subsection{Interaction-induced instability of polariton spin dynamics}

The linear analysis of kinetic equation (5) in the main text shows that the polariton spin system becomes unstable at $\theta=\pi/2$ ($\bm S_0 \parallel \bm\Omega_L$ is in the structure plane) provided that
\begin{equation}
\alpha _{1}S_{0}\Omega _{L}>\Omega _{L}^{2}+\frac{1}{\tau _{c}^{2}},
\label{instability}
\end{equation}%
see Eq. (7) of the main text. In the linear regime the excitations are grow exponentially (in the unstable regime) with the increment
\begin{equation}
\label{increment}
\lambda = \sqrt{\alpha _{1}S_{0}\Omega _{L} - \Omega_L^2} -1/\tau _{c}
\end{equation}
In order to analyze the instability in more detail we consider the simplest possible case where polariton generation and dissipation are absent and the spin dynamics is described by the following equation
\begin{equation}
\frac{\partial \bm S}{\partial t} + \bm S\times \bm \Omega^{(\rm eff)} = 0,
\label{dynamics}
\end{equation}
with 
\begin{equation}
\bm\Omega _{\bm k}^{(\mathrm{eff)}}=\bm\Omega _{L}+\alpha _{1}\sum_{\bm %
k^{\prime }}S_{\bm k^{\prime },z}\bm e_{z}, \label{field}
\end{equation}%
see Eqs. (1) and (2) of the main text.

\begin{figure}[h]
\includegraphics[width=\linewidth]{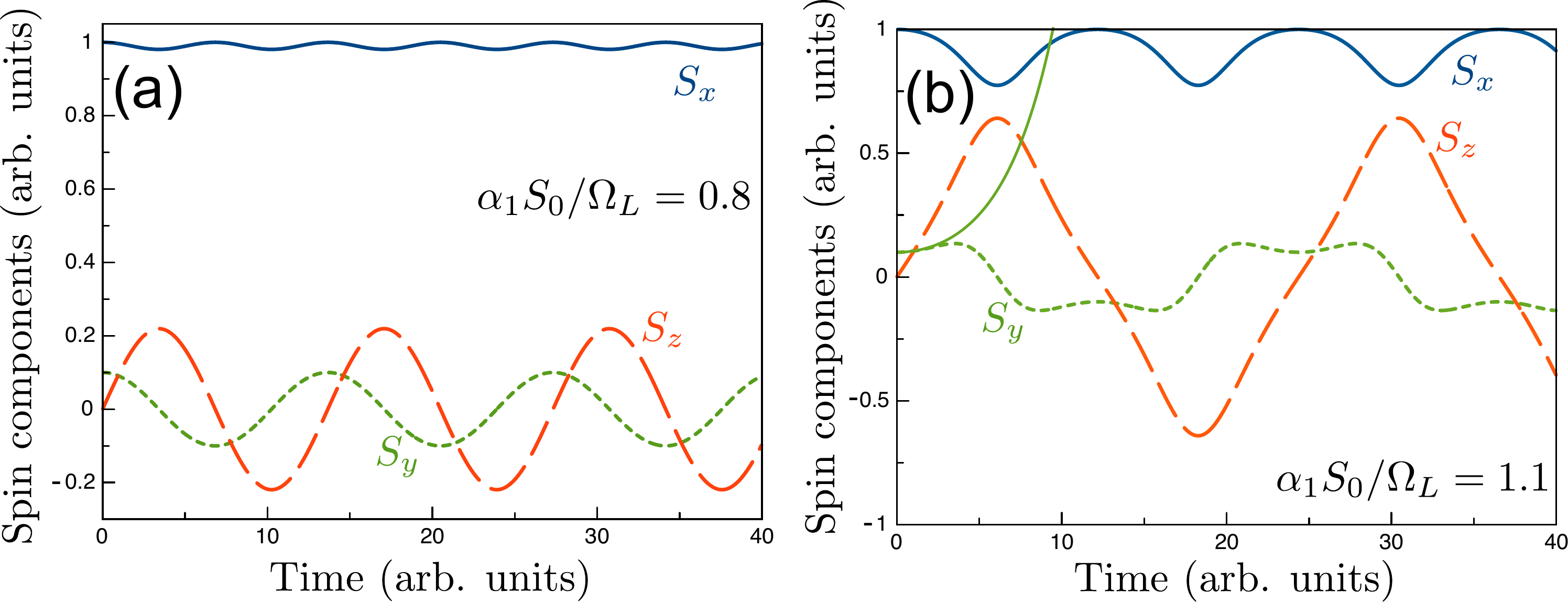}
\caption{Panels (a) and (b) show temporal dynamics of the total spin components nonlinear Eqs.~\eqref{dynamics}, \eqref{field} for homogeneous case for $\bm \Omega_L \parallel \bm S_0 ||x$ (at $t=0$) and small $\delta s_y(0)=S_0/10$. (a) corresponds to stable and (b) to unstable regimes. Spin components are marked at each curve, thin solid line shows exponentially growing solution calculated after Eq.~\eqref{increment}. }
\label{fig:supp:1}
\end{figure}

 Examples of temporal dynamics of the total polariton pseudospin in the stable and unstable regimes calculated numerically after Eq.~\eqref{dynamics} are shown in Fig.~\ref{fig:supp:1}, panels (a) and (b), respectively. It is seen that the instability of small spin fluctuations can result in the nonlinear oscillations of spin polarization similar to those discussed in Refs.~\cite{PhysRevB.80.1953091,Kammann:2012uq1} [Fig.~\ref{fig:supp:1}(b)]. The detailed analysis of the final state of the system and its stability with allowance for dissipative processes is beyond the scope of the present paper.

\subsection{Spin precession caused by TE-TM splitting}

Here we demonstrate that the TE-TM splitting of the polariton states can also result in the weakly damped spin resonance and diffusive spin modes. We consider symmetric cavity with the TE-TM splitting $\bm \Omega_{\bm k}$ in the form
\begin{equation}
\bm \Omega_{\bm k} = \Omega(k)[\cos{\varphi}, \sin{\varphi},0],
\end{equation}
where $\varphi$ is the angle between $\bm k$ and $x$-axis in the structure plane, $\Omega(k)$ is the amplitude of the TE-TM splitting. For simplicity we consider monoenergetic polaritons with the energy $\varepsilon_0 \equiv \varepsilon(k_0)$, introduce $\Omega_0 = \Omega(k_0)$, and assume that both $\Omega_0\tau_c,\Omega_0\tau\gg 1$. We focus on dynamics of spin $z$ component. For $\bm q=0$ we obtain (note that $S_z$ is isotropic function of $\varphi$, while $S_x$ and $S_y$ are strongly anisotropic and their angular averages are $0$):
\begin{subequations}
\label{TETM:hom}
\begin{align}
\left(-\mathrm i \omega + \frac{1}{\tau_c}\right) S_{k,z} + S_{\bm k,x} \Omega_{0,y} - S_{\bm k,y} \Omega_{0,x} &= 0,\label{z}\\
\left(-\mathrm i \omega + \frac{1}{\tau_c}  +\frac{1}{\tau} \right) S_{\bm k,x} - S_{k,z} \Omega_{0,y}&=0, \label{x}\\
\left(-\mathrm i \omega + \frac{1}{\tau_c}  +\frac{1}{\tau} \right) S_{\bm k,y} + S_{k,z} \Omega_{0,x} &=0. \label{y}
\end{align}
\end{subequations}
Solution of Eqs.~\eqref{TETM:hom} yields the frequency of homogeneous spin $z$ component oscillations
\begin{equation}
\omega = \Omega_0 \sqrt{1-[2\Omega_0\tau]^{-2}} - \frac{\mathrm i}{\tau_c} - \frac{\mathrm i}{2\tau}.
\end{equation}
In the relevant limiting case $\Omega(k)\tau \gg 1$ spin $z$ component demonstrates oscillations with the frequency $\Omega(k)$ similarly to the oscillations of spin in high-mobility electron gas~\cite{gridnev011,brand021,glazov20071}. Note that for $\Omega(k)\tau \ll 1$ the frequency is purely imaginary, $\omega = - \mathrm i \Omega^2_0\tau - \mathrm i/\tau_c$.

The spectrum of inhomogeneous spin excitations can be conveniently found in the limit of $qv_0\tau \ll 1$, in which case the gradient term $\propto (\bm q \bm v_{\bm k})$, Eqs. (1) and (4) of the main text, can be taken into account by perturbation theory. After some algebra we obtain for monoenergetic particles in the limit of $\Omega_0 \tau_c, \Omega_0\tau\gg 1$, $qv_0\tau\ll 1$:
\begin{equation}
\label{disper}
\omega(q) = \Omega_0 - \frac{\mathrm i }{\tau_c} -\frac{\mathrm i}{2\tau} - \mathrm i (qv_0)^2\tau.
\end{equation}
Unlike the spin waves predicted in the main text, Eq. (10), here the spatial inhomogeneity results in the diffusive damping of the spin precession mode. The analysis of the interplay between interactions and TE-TM splitting is beyond the present paper.

\subsection{Spatial correlations of polariton spins}

Spin waves describe the time-space correlations of polariton spins and can be addressed in the 
two-beam spin noise spectroscopy technique (for reviews on the spin noise spectroscopy see, e.g., Refs.~~\cite{Zapasskii:131,Oestreich:rev1}, temporal and spatial fluctuations of spin density were studied for electrons for the first time in Refs.~\cite{ivchenko73fluct_eng1,ustinov1975scattering1}, respectively). The correlation function of polariton spin fluctuations can be expressed via the spin waves spectrum $\omega(q)$ as
\begin{equation}
\label{space}
\langle \delta S_z(\bm r, t) \delta S_z(0,0)\rangle \propto \int \frac{\mathrm d\omega}{2\pi} \sum_{\bm q} \frac{\exp{(\mathrm i \bm q \bm r - \mathrm i \omega t)}}{\omega - \omega(q)}.
\end{equation}
In the long-wavelength limit [see Eqs. (10) and (11) of the main text]
\begin{equation}
\label{dispersion}
\omega (q) = \Omega - \mathcal A q^2 - \mathrm i /\tau_c,
\end{equation}
where $\Omega$ and $\mathcal A$ are the parameters depending on the effective field, $\bm S_0$ and interactions. Note that diffusion of particles yields imaginary part $\mathcal A''$ of $\mathcal A = \mathcal A' + \mathrm i \mathcal A''$ in addition to its real part $\mathcal A'$ determined by interactions. Making use of Eq.~\eqref{dispersion} we arrive from Eq.~\eqref{space} to 
\begin{equation}
\label{corr:space}
\langle \delta S_z(r, t) \delta S_z(0,0)\rangle  = \Im\left\{\frac{\exp{\left[\mathrm i \left(\frac{r^2}{4\mathcal A t} - \Omega t \right)\right]}}{4\pi \mathcal A t} \right\} \mathrm e^{-t/\tau_c}.
\end{equation}

\begin{figure}[tb]
\includegraphics[width=\linewidth]{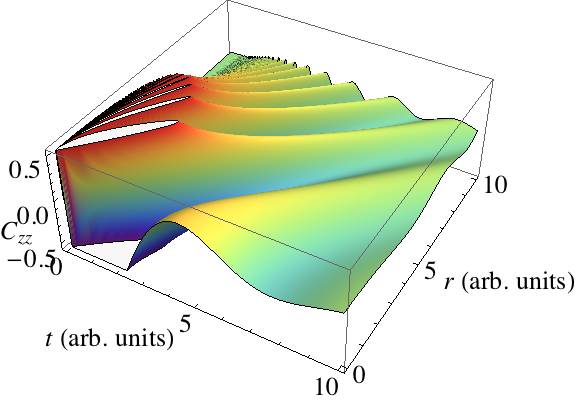}
\caption{Correlator of polariton spins $C_{zz} =\langle \delta S_z(r, t) \delta S_z(0,0)\rangle$  calculated after Eq.~\eqref{corr:space}. The calculation parameters are as follows: $\Omega = 1$, $\tau_c=10$, $\mathcal A'= 0.25$, $\mathcal A''=0.01$ (coordinates and time are given in dimensionless units).}
\label{fig:supp:2}
\end{figure}

The spin correlation function is presented in Fig.~\ref{fig:supp:2}. The oscillations of the correlation function as a function of coordinate and time are clearly seen demonstrating wave-like propagation of spin excitations.


\begin{thebibliography}{99}
\bibitem{gen} {Results for the general case will be presented elsewhere.}
\bibitem{Zeeman} The Zeeman effect caused by the in-plane components of
external field is negligible since it mixes dark and bright excitonic states.
\bibitem{TETM} The TE-TM splitting itself results in the non-trivial spin dynamics of polaritons~\cite{gen,suppl}. It can be also suppressed by design of microcavity~\cite{panz}.
\end{thebibliography}

\end{document}